\documentclass[prd, twocolumn, nofootinbib, floatfix]{revtex4}

\usepackage{amsmath}
\usepackage{amsbsy}

\input epsf
\usepackage{amsmath,amssymb,subfigure}
\usepackage{graphicx}
\usepackage{epsfig}
\usepackage{color}
\usepackage{ulem}
\usepackage{epstopdf}

\newcommand{\ls}{\mathrel{\raise0.27ex\hbox{$<$}\kern-0.70em \lower0.71ex\hbox{{
$\scriptstyle \sim$}}}}

\begin{document}

\title{Oscillations in the Primordial Bispectrum: Mode Expansion}

\author{P. Daniel Meerburg}
\affiliation{Astronomical Institute ``Anton Pannekoek", University of Amsterdam, Amsterdam 1098 SJ, The Netherlands; p.d.meerburg@uva.nl}

\date{\today}

\begin{abstract} 
We consider the presence of oscillations in the primordial bispectrum, inspired by three different cosmological models; features in the primordial potential, resonant type non-Gaussianities and deviation from the standard Bunch Davies vacuum. In order to put constraints on their bispectra, a logical first step is to put these into factorized form which can be achieved via the recently proposed method of polynomial basis expansion on the tetrahedral domain. We investigate the viability of such an expansion for the oscillatory bispectra and find that one needs an increasing number of orthonormal mode functions to achieve significant correlation between the expansion and the original spectrum as a function of their frequency. To reduce the number of modes required, we propose a basis consisting of Fourier functions orthonormalized on the tetrahedral domain. We show that the use of Fourier mode functions instead of polynomial mode functions can lead to the necessary factorizability with the use of only $1/5$ of the total number of modes required to reconstruct the bispectra with polynomial mode functions. Moreover, from an observational perspective, the expansion has unique signatures depending on the orientation of the oscillation due to a resonance effect between the mode functions and the original spectrum. This effect opens the possibility to extract information about both the frequency of the bispectrum as well as its shape while considering only a limited number of modes. The resonance effect is independent of the phase of the reconstructed bispectrum suggesting Fourier mode extraction could be an efficient way to detect oscillatory bispectra in the data. 
 \end{abstract}



\maketitle

\section{Introduction}

In recent years it has become evident that determining the precise physics of inflation requires the observation of higher order correlation functions beyond the power spectrum \cite{Komatsu:2009kd}. These correlation functions can 
be obtained from the Cosmic Microwave Background (CMB) \cite{WMAP5,Komatsu:2010fb,Senatore:2009gt,Smidt:2010ra,Smith:2009jr} and Large Scale Structure (LSS) \cite{Xia:2010yu,Verde:2010wp}, but recently \cite{Cooray:2006km,Pillepich:2006fj,Cooray:2008eb}, it has been shown that in principle 21-cm observations of the early universe can also be used to measure n-point statistics. Because higher order correlation functions introduce more free parameters they can 
be used to constrain more complex models of inflation, since an increased set of parameters will allow for a unique fitting of the model to the observed data \cite{Komatsu:2009kd}. However, both due to
computational and observational limitations, only the bispectrum has been reasonably investigated. For the detection of higher order correlations we will have to wait for more advanced data sets, such
as Planck and improved analysis methods, although preliminary attempts have been made \cite{Smidt:2010ra,Smidt:2010sv,Regan:2010cn}. Even the detection of the bispectrum is not optimal, as a bispectrum would at least be a continuous three parameter observable but thus far only constraints have been set on
limiting cases, in which 2 of the parameters are fixed and the third one is measured for a predetermined triangular configuration. The limiting cases (shapes) are known as the local, equilateral and orthogonal (and in the context of limiting triangular configurations; enfolded) non-Gaussian features. 
Precisely these features have been chosen, as it has been shown theoretically that most models of inflation produce non-Gaussianities that fall in one of these three classes (for recent reviews see \cite{Chen:2010xk,Bartolo:2010qu,Komatsu:2010hc}). 

When constraining non-Gaussianities using the bispectrum, it has been a prerequisite that the comoving momentum dependence should be factorizable; the bispectrum should be separable into a product of functions of one variable, each variable being one of the three comoving momenta making up the connected correlation triangle. Foremost, this requirement is set because of computational limitations that would render the analysis intractable if a given primordial bispectrum is not of the factorized form. The number integrals and sums one has to perform when computing an unfactorized bispectrum scale with the number of pixels as $N^{5/2}$, while for factorizable shapes this reduces by one factor of $N$ \cite{Wang:1999vf}. Although one integral can be computed fairly quickly the number of pixels ($\mathcal{O}(10^6)$ for WMAP and $\mathcal{O}(10^7)$ for Planck) is large and one factor of $N$ can make all the difference. The constrained bispectra, local, equilateral and orthogonal, have thus far been factorized templates. In case of equilateral \cite{Creminelli:2005hu} and orthogonal \cite{Senatore:2009gt} these have been constructed via approximation of a predicted signal, in the local case, the template is a direct representation of the theory \cite{Gangui:1993tt,Komatsu:2001rj,Maldacena:2002vr}. For a particular type of bispectrum to be constrained, it is necessary to construct a factorized template that `matches' the bispectrum. Until recent, there was no given prescription how to factorize a given theoretical bispectrum. In \cite{Fergusson,Fergusson:2009nv} it was shown that factorizability can be achieved in both comoving momentum and multipole space by expanding the bispectrum in mode functions that are orthogonal on the domain of the bispectrum dictated by triangle constraints. The purpose of this factorization is to be able to quickly compute the full CMB bispectrum ($B_{l_1 l_2 l_3}$) and generate CMB maps with a arbitrary primordial statistics (up to the trispectrum \cite{Regan:2010cn}) which are used to determine the variance of the (statistical) estimator. In the same paper, it was also shown that one can efficiently extract information about non-Gaussianity in the observed CMB by measuring the weight of each mode in the data and comparing this to theoretical predictions\footnote{During the finalization of this paper, the same group published a paper \cite{Fergusson:2010dm} in which many non-factorizable non-Gaussian shapes have been constrained using the WMAP 5 year data and the method of mode expansion}. 

In this paper we investigate how well this mode expansion works for a class of bispectra that contain (a large number of) oscillations. The reason to be interested in such features is that a number of theoretical models \cite{Bean:2008na,Chen:2008wn,Meerburg2009a,Meerburg2009b,Flauger:2009ab} predict oscillations in the bispectrum and in order to be able to constrain such models, a plausible first step is to factorize these bispectra.  As it is, oscillations can be considered as an extra, distinguishable, degree of freedom within the bispectrum which could result in narrowing down the number of potential scenarios of inflation.

We introduce three different cosmological scenarios in which oscillations in the bispectrum can appear. We will briefly discuss the theory behind these models and show to what extend these would be distinguishable from one another in the data in section \ref{Oscillations}. Two out of three bispectra can have significant correlation and it could be difficult to discriminate between such models in future surveys. We will discuss the method of polynomial expansion in order to rewrite the primordial bispectra in factorized/separable form in section \ref{Mode_expansion}. As expected, the number of modes required in the expansion grows along with the frequency of the theoretical spectra. In section \ref{Powermodes} we show how fast polynomial expansion would yield a reasonable reconstruction of the given bispectra predicted by the three cosmological scenarios. Subsequently we will investigate another set of modes that can lead to a separable expansion of the theoretical bispectrum in section \ref{Fouriermodes}. These modes are based on the sine and cosine and the resulting set of orthonormal functions can be considered a Fourier-type basis on the tetrahedral domain.  After detailing the construction of this set of orthonormal mode functions, we will compare the number of modes required to achieve comparable correlation with the polynomial mode expansion . It turns out that this number is reduced significantly and as such Fourier expansion can be considered a reasonable alternative to expand oscillatory spectra. For larger frequencies both Fourier and polynomial mode expansion become inefficient. Fortunately, for various oscillatory signals only a limited number of modes contribute significantly in the reconstruction of the original spectrum. This has several consequences for the viability of Fourier mode expansion as well as possible observational advantages compared to polynomial modes, which will be discussed in section \ref{discussion}. In these class of models, just as the frequency, the phase can be considered a free parameter of the theory. In a polynomial mode expansion, different phases can result in significantly different expansions. In a Fourier mode expansion the phase is taken care of much more naturally. Effectively the phase can be absorbed into the weights of the expansion, and as such have minimal effect on the overall expansion. Consequently, we will see that the norm of the mode expansion coefficients will be very similar for each phase making Fourier expansion much more elegant and suitable for these type of spectra. We conclude this paper in section \ref{conclusion}. 

\section{Oscillations in Primordial Bispectra}\label{Oscillations}

In this section we will briefly discuss 3 distinct possibilities that can produce non-Gaussianities that have an oscillatory component. Two of these examples have an exact solution, while a third has only been solved numerically and we will use an approximate form.  In the following paragraphs we will describe the physics behind these models and quote their theoretically predicted primordial bispectra. In addition we investigate how well these bispectra can be distinguished from one another by computing their correlation, which will be defined shortly. Since all these bispectra have poor overlap with existing spectra, there exists substantial room for improvement, which we could achieve by approximating these shapes via mode expansion. This will be the topic of the next section.

For completeness, let us introduce (standard) notation. The primordial bispectrum is given by 
\begin{eqnarray}
\langle \zeta_{\vec{k}_1}  \zeta_{\vec{k}_2}  \zeta_{\vec{k}_3} \rangle&=& (2 \pi)^7 f_{NL}\Delta^2 \delta^K\left(\sum_{i=1}^3 k_i\right) F(k_1,k_2,k_3),\nonumber\\
\end{eqnarray}
where $\zeta$ is the gauge invariant curvature perturbation ($\zeta = -H\delta \phi /\dot{\phi}_0$) which is constant after horizon exit, $\Delta$ is the amplitude of the primordial power spectrum (i.e. for single field slow-roll $\Delta=H^2/8\pi \epsilon$, where $H$ is the Hubble rate at the end of inflation and $\epsilon$ the slow-roll parameter) and $F(k_1,k_2,k_3)$ is the shape of the bispectrum. We will also make use of $S\equiv k_1^2k_2^2k_3^2 F$. In the following we will discuss the shapes of the bispectra and quote theoretically predicted ranges of their associated $f_{NL}$. We would like to refer to the literature for a detailed examination of the theoretically predicted values of $f_{NL} $\cite{Chen:2008wn,Meerburg2009a,Meerburg2009b,Flauger:2009ab} in various theoretical contexts.

\subsection{Features in the Potential}\label{sharpfeatures}

Sharp features in the potential can temporarily break  slow-roll and produce large non-Gaussianities  \cite{Chen:2006xjb,Chen:2008wn}. As long as the system relaxes within several Hubble times, inflation can still lead to a significant amount of e-folds to solve the standard cosmological problems. The motivation for these type of features is two-fold. First, there are hints of glitches in the primordial power spectrum that could be cross-checked using the bispectrum \cite{Covi:2006ci}.  A second motivation is theoretical in nature. In certain brane inflation models the effective 4-dimensional potential displays sharp features (see \cite{Chen:2010xk} and references therein). 

One of the possible sharp features is a step in the potential, which can be parameterized as
\begin{eqnarray}
V(\phi) &  = &\frac{1}{2}m^2 \phi^2 \left[1+c\tanh\left(\frac{\phi-\phi_s}{d}\right)\right],
\end{eqnarray}
where $c$, $d$ and $\phi_s$ respectively determines the height, width and location of the feature. 
 
The resulting bispectrum can only be computed numerically. The authors of \cite{Chen:2008wn} have proposed an approximate analytic form

\begin{eqnarray}
F_{Feat}&\simeq& \frac{\sin (k_t/k_*+\delta)}{k_1^2 k_2^2 k_3^2}.
\end{eqnarray}
The approximation can in principle be improved \cite{Chen:2010xk} by multiplying by an `envelope' function, but such improvement would not gain us any more useful insight required for the analysis in this paper and we will therefore omit it.  Here $k_*$ is related the location of the feature in the potential $\phi_s$. Evidence for features in the power spectrum around $l\sim 30$ have been put forward in \cite{Covi:2006ci}. It was shown that the inclusion of features in the primordial potential could improve the $\chi^2$ best-fit.  Such a feature would approximately correspond to $k_* = 30/\eta_0 \sim0.002\mathrm{Mpc}^{-1}$. This relation also indicates that the smaller the scale at which the feature appears the larger the associated wavelength. Roughly the wavelength corresponds to the location of the feature, e.g. for a feature at $l=30$ the wavelength $\delta l\sim30$ \cite{Bean:2008na}. Here we do not necessarily relate to an observed feature at a specific value in multipole space since features that lead to non-vanishing bispectra can still be present with minimal consequences for the observable power spectrum. The quantities we will compute in the remainder of this paper are mostly integrals that run over the domain of comoving momentum space between $k_{min}\leq k\leq k_{max}$. It is therefore convenient to choose our reference scale $k_{max}\sim10^{-1}\mathrm{Mpc}^{-1}$, the smallest observable scale in the data, in order to be able to compare the frequencies in the various models. We then define $x_1=k_1/k_{max}$, $x_2=k_2/k_{max}$, $x_3=k_3/k_{max}$, $x_t=k_t/k_{max}$ and rewrite the shape of this bispectrum as
\begin{eqnarray}
F_{Feat}&=& k_{max}^{-6}\frac{sin (\omega_{f} x_t +\delta)}{x_1^2 x_2^2 x_3^2},
\label{eq:feat_bispectrum}
\end{eqnarray}
with $\omega_{f} = k_{max}/k_*$. For a feature at $k_*=0.002\mathrm{Mpc}^{-1}$ we therefore find $\omega_f \sim 50$. Note that $\omega=50$ can be considered an upper limit in allowable frequencies due to features in the potential. For features at smaller scales the frequency will be smaller. This bispectrum with a frequency of $\omega_f=50$ is shown in the bottom of figure \ref{fig:3Dbispectra}.

The amplitude of this type of non-Gaussianity is governed by the width and the depth of the feature in the potential 
\begin{eqnarray}
f_{NL}^{feat} &\sim& \frac{7 c^{1/2}}{d \epsilon},
\end{eqnarray} 
which for a feature at $l\sim 30$ would imply $f_{NL}^{feat} \sim \mathcal{O}(10)$ \cite{Covi:2006ci}.

\subsection{Resonant non-Gaussianity}\label{resonant}

This type of non-Gaussianity is a result of a periodic feature in the inflaton potential as apposed to a sharp feature explored in the previous example. These features will cause oscillations in the coupling(s) of the interaction terms of the inflaton field. Resonance occurs when an oscillatory mode well within the horizon grows during inflation until its frequency hits the same frequency as those of the couplings.  So as long as $\omega>H$ resonance will occur at some point within the inflationary history of the mode. This resonance can result in a large contribution to the three point correlation function  \cite{Chen:2008wn}. 

In a general scenario, with an oscillatory potential we obtain an expression for the bispectrum of the form  \cite{Chen:2008wn,Flauger:2009ab}
\begin{eqnarray}
F_{res}&=& \frac{1}{k_1^2 k_2^2 k_3^2}\Biggl( \sin (C \ln (k_t/k_*)) \nonumber\\
&&\left.+ C^{-1} \cos (C \ln (k_t/k_*))\sum_{i\neq j}\frac{k_i}{k_j}\right)
\end{eqnarray}
Here $C$ is related to the frequency as $C=\omega/H$ with $H$ the Hubble rate during inflation (which is approximately constant) and $k_*$ introduces a phase. One can also compute the general expected amplitude of non-Gaussianity which is related to the frequency as 
\begin{eqnarray}
f_{NL}^{res}&\sim& \frac{\sqrt{\pi}}{2\sqrt{8}}\frac{\omega^{1/2}\dot{\eta}_A}{H^{3/2}}.
\end{eqnarray}
Here $\eta_A$ represents the amplitude of the oscillatory component of the couplings. 

Physically such features might be realized in terms of brane inflation \cite{Bean:2008na} where the periodic feature comes from a duality cascade in the warped throat, as well as axion-monodromy inflation where the periodic feature is a result of instanton effects \cite{Flauger:2009ab,Flauger:2010ja}. As an example let us consider the latter. Axion inflation is well embedded in string theory and represents a favorable candidate for inflation if the observed tensor modes are relatively large ($r\sim0.07$). Such a scenario implies inflation occurred at energies close to the GUT scale and would indicate that we require the knowledge of the UV completion.

\begin{figure*}
   \centering
   \includegraphics[scale=.60]{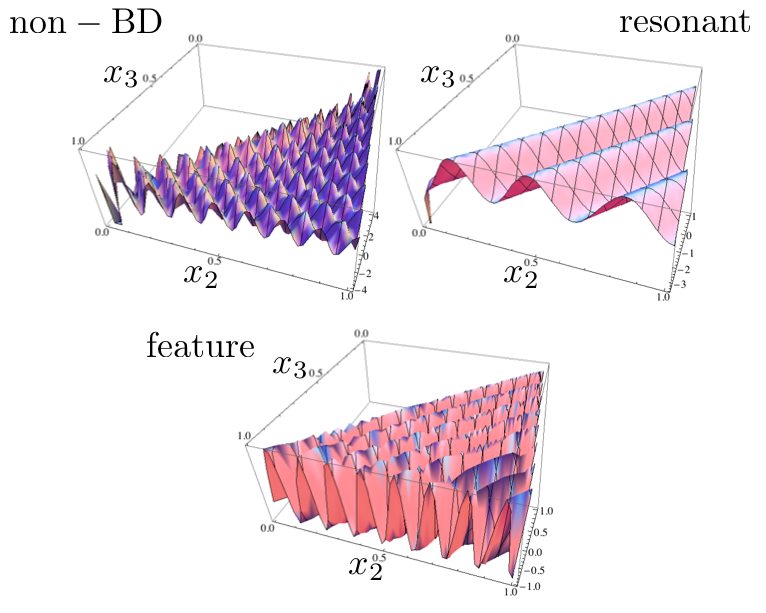} 
   \caption{3 examples of oscillating bispectra. We have set $\omega_v=\omega_f=\omega_r=50$. The pivot scale in eq. \protect\eqref{eq:res_bispectrum} is set to $k_* = 0.002\mathrm{Mpc}^{-1}$ and $x_1=1\rightarrow k_1=k_{max}$. In addition $\delta =0$ in both the modified initial state and feature scenario. The non-BD bispectrum contains the most features and, not clear from this image, the number of features (effective frequency) increases rapidly for smaller values of $x_1$ making this bispectrum particularly hard to reconstruct using mode expansion.} 
    \label{fig:3Dbispectra}
\end{figure*}

The axion potential is given by
\begin{eqnarray}
V(\phi)&=&V_0(\phi)+\Lambda^4 \cos \phi /f.
\label{eq:potential}
\end{eqnarray}
The parameter $f$ represents the axion decay parameter. The range of $f$ which would generate observable non-Gaussianities and is still consistent with observations of the power spectrum is given by $10^{-4}\lesssim f\lesssim6\times 10^{-3}$ \cite{Flauger:2009ab,Flauger:2010ja} . The lower bound is set by the requirement that the period of the oscillation should be larger than $\Delta l \sim 1$ for $l\equiv 200$. For a linear zero order potential the resulting bispectrum is then given by
\begin{eqnarray}
F_{res}&=& \frac{k_{max}^{-6}}{x_1^2 x_2^2 x_3^2}\Biggl( \sin (\omega_{r} \ln x_t+\gamma_1) \nonumber\\
&&\left.+\omega^{-1}_r \cos (\omega_{r} \ln x_t+\gamma_1)\sum_{i\neq j}\frac{x_i}{x_j}\right),
\label{eq:res_bispectrum}
\end{eqnarray}
with $\omega_{r} = (f\phi_*)^{-1}$ and $\gamma_1 = \omega_r \ln k_{max}/k_*$. Here $k_*$ is  pivot scale ($k_*=0.002 \mathrm{Mpc}^{-1}$), and $\phi_*$ is the value of the inflaton field when the pivot scale exits the horizon and is of order 10 ($M_p$).  Given the range for $f\phi_*$ the frequency of the oscillations in the bispectrum lie within $ 20\lesssim \omega_r \lesssim 10^3$.  A plot of this shape is shown in the top right figure \ref{fig:3Dbispectra}.

The amplitude of the axion bispectrum (for a linear potential) is given by
\begin{eqnarray}
f_{NL}^{res} &=& \frac{3 \sqrt{2\pi}b}{8(f\phi_*)^{3/2}}.
\end{eqnarray}
The amplitude is therefore proportional to a power of the frequency. For a linear potential $b=\Lambda^4/(\mu^3 f)$, where $\mu\sim 6\times 10^{-4}$ is fixed by COBE normalization. From observations of the power spectrum one can constrain $b f<10^{-4}$ \cite{Flauger:2009ab} and therefore
\begin{eqnarray}
f_{NL}^{res} &\sim& 10^{-3} \omega_r^{5/2},
\end{eqnarray}
allowing $\mathcal{O}(1)\leq f_{NL}^{res}\leq \mathcal{O}(10^4)$.

\subsection{Initial State Modifications}\label{nonBD}

Since inflation is an effective field theory in a curved background, choosing an appropriate vacuum state is by no means evident\footnote{For a in depth discussion on deviations from a BD vacuum see for example \cite{Meerburg2009b,Schalm:2004xg} and \cite{Chen:2010xk} section 6.4}.  In general the initial or vacuum state is chosen to be equivalent to the free field vacuum state in flat Minkowski space, know as the Bunch Davies (BD) vacuum. Although it seems that possible corrections to this assumption are constrained to be small (from general observation of the power spectrum \cite{Flauger:2009ab,Pahud:2008ae,Easther:2005yr}  and backreaction constraints \cite{Schalm:2004xg,Holman:2007na}), it has been shown that small corrections in the BD state can result in rather large non-Gaussian effects \cite{Chen:2006nt, Holman:2007na, Meerburg2009a, Meerburg2009b}. Using the currently available bounds on non-Gaussianity from CMB data, deviations from a pure Bunch Davies stae have been constraint even further, although these constraints strongly depend on the inflationary model. However, there exist significant room for improvement as non-Gaussianities from these modifications are highly oscillatory and therefore the derived constraints are relatively poor since they depend on the correlation with measured smooth bispectra. 

A number of different scenarios have been considered in which initial state modifications were investigated. Here we will not discuss all of these, although the results can differ significantly \cite{Meerburg2009b}. Such differences make it difficult to make robust predictions, it seems inevitable however that once you introduce a effective field theory cutoff, oscillations appear in both the power and  bispectrum. We will consider one example that represents a large class of models with a non-canonical effective field theory action, which already drives large non-Gaussianities to start with. This particular class has a speed of sound $c_s<1$, such that perturbations in the medium propagate slow compared to the growth of the causal horizon. The leading order shape of the resulting bispectrum is given by \cite{Meerburg2009b}
\begin{eqnarray}
F_{BD}&=& \frac{c_s\eta_0}{ k_1k_2k_3} \sum_j \left(\frac{1}{2} \frac{\cos (\tilde{k}_j c_s \eta_0 +\delta)}{\tilde{k}_j c_s \eta_0}-\right.\\
& & \left.\frac{\sin (\tilde{k}_j c_s \eta_0 +\delta)}{(\tilde{k}_j c_s \eta_0)^2}+ \frac{\cos \delta -\cos (\tilde{k}_j c_s \eta_0 +\delta) }{(\tilde{k}_j c_s \eta_0)^3} \right).\nonumber
\end{eqnarray}
Here $\tilde{k}_j=k_t-2k_j$. In \cite{Meerburg2009a} it was assumed that there exists a fixed physical cutoff hyper-surface $\eta_0$ that is scale dependent such that the overall momentum dependence of the bispectrum becomes scale invariant. Such a choice is known as the New Physics Hypersurface (NPH), as apposed to Boundary Effective Field Theory (BEFT) approach in which the cutoff is  time dependent \cite{Greene:2005aj}. The subtlety is that the cutoff appears due to the presence of a non-BD state in each direction in comoving momentum space. Consequently, $\eta_0(k_i)$ will depend on the $k_i$ direction the BD vacuum has been perturbed in. This direction is set by the direction in which $k_i$ picks up a minus sign due to the Bunch Davies vacuum perturbation as explained in \cite{Meerburg2009a}. One could allow for scale invariance breaking and consider BEFT, however there are some suggestions \cite{Sefusatti:2009xu} that such large scale invariance should have been observed already. We can rewrite the bispectrum as
\begin{eqnarray}
F_{nBD} &=& \frac{\omega_{v}k_{max}^{-6}}{ x_1 x_2 x_3} \sum _j\frac{1}{x_j^3}\left(\frac{1}{2}\frac{\cos \left(\omega_{v} \frac{x_{j+1}+x_{j+2}}{x_{j}}+\gamma_2\right)}{\omega_v \left(\frac{x_{j+1}+x_{j+2}}{x_j}-1\right)} \right.\nonumber\\
&&- \frac{\sin \omega_{v} \left(\omega_{v} \frac{x_{j+1}+x_{j+2}}{x_{j}}+\gamma_2\right)}{\omega_v^2 \left(\frac{x_{j+1}+x_{j+2}}{x_j}-1\right)^2}\nonumber\\
&&\left. \frac{\cos \delta-  \cos \left(\omega_{v} \frac{x_{j+1}+x_{j+2}}{x_{j}}+\gamma_2\right) }{\omega_v^3 \left(\frac{x_{j+1}+x_{j+2}}{x_j}-1\right)^3}\right), 
\label{eq:nbd_bispectrum}
\end{eqnarray}
where $\gamma_2= \delta -\omega_v$ and $\omega_{v} = k \eta_0 c_s=(k /a_0)/(H/c_s)$ or the ratio between the largest physical momentum scale and the Hubble radius at time $\eta_0$ which can be as large as $10^3$ \cite{Meerburg2009a, Meerburg2009b, Holman:2007na}. Note that from this expression it seem that $x_j=x_{j+1}+x_{j+2}$ represents a singular line (the enfolded limit). However, one can show that all infinities are cancelled against each other and the the expression is finite and vansihing\footnote{This limit is on the enfolded line $x_j=x_{j+1}+x_{j+2}$ within the sum. Outside the sum, this expression is non-zero but finite. For example $x_1\rightarrow x_2+x_3$ gives:
\begin{eqnarray}
\frac{1}{8 x_2^4 x_3^4 \omega ^2}\left((x_2+x_3)^3-z \left((x_2+x_3)^2-2 x_2^2 \omega ^2\right) \cos \left(\frac{2 x_2 \omega}{x_2+x_3}\right)\right.\nonumber\\
-x_2 \left(\left((x_2+x_3)^2-2 x_3^2 \omega ^2\right) \cos \left(\frac{2 x_3 \omega}{x_2+x_3}\right)\right.\nonumber\\
+\left.\left.2 x_3 (x_2+x_3) \omega  \left(\sin \left(\frac{2 x_2 \omega }{x_2+x_3}\right)+\sin \left(\frac{2 x_3 \omega }{x_2+x_3}\right)\right)\right)\right).\nonumber\end{eqnarray} 
}.When computing quantities numerically, such as the correlator in section $3$, these apparent singularities can be hard to handle and we need to be aware of these. We have plotted this shape in the top left figure \ref{fig:3Dbispectra}. 

The amplitude of the non-BD bispectrum is a function of the frequency and the Bogoliubov parameter quantifying the deformation away from the BD state.The way this bispectrum was computed, considered a Bogolyubov correction of linear order $\beta$ and small speed of sound $c_s$. In this particular scenario, $f_{NL}$ is roughly given by
\begin{eqnarray}
f_{NL}^{nBD}&\sim& \frac{1}{c_s^2} \omega_v^3 \beta.
\end{eqnarray}
From backreaction and power spectrum constraints $\beta \lesssim 10^{-2}$, which could still allow observable levels of non-Gaussianity.

\subsection{Distinguishability}\label{distinguishability}

Although the presented theoretical bispectra have different characteristics, we would like to get an indication how well these could be discriminated. For instance, it seems obvious that the similarity between the feature bispectrum and the resonant bispectrum could lead to significant confusion when actually traced in the data. In order to do so, we want to measure the distinguishability of these shapes, which is usually quantified using the amount of overlap or correlation between two shapes. One can define a inner product between two shapes
\begin{eqnarray}
F_{X}\star F_{Y} & \equiv & \int_{\Delta_{k}}dk_{1}dk_{2}dk_{3}k_{1}^{4}k_{2}^{4}k_{3}^{4}w_k F_{X}F_{Y}\nonumber\\
&=& \int_{\Delta_{k}}dk_{1}dk_{2}dk_{3} w_k S_{X}S_{Y}. 
\label{eq:dotproduct}
\end{eqnarray}
The correlation between two shapes $F_X$ and $F_Y$ is then defined as 
\begin{eqnarray}
\mathrm{C}(F_{X},F_{Y}) & \equiv & \frac{F_{X}\star F_{Y}}{(F_{X}\star F_{X})^{1/2}(F_{Y}\star F_{Y})^{1/2}} .
\label{eq:correlator}
\end{eqnarray}
Here $w_k$ is a weight function, which was chosen as $w_k=1/k_t$ in \cite{Fergusson:2009nv} to increase resemblance with the Fisher matrix (correlation) found in multipole space. The integral runs over the `tetrahedral' domain, which is bounded by the following triangle constraints
\begin{eqnarray}
k_a\leq k_b+k_c\;\mathrm{for}\;k_a\geq k_b,k_c\nonumber\\
k_a,k_b,k_c\leq k_{max}\nonumber,
\end{eqnarray}
where $a,b,c=\{1,2,3\}$, $a\neq b\neq c$.

Before we compute the correlation between the shapes, let us perform a quick qualitative analysis in order to get an indication of what to expect. First of all, note that the shape coming from initial state modifications (eq. \eqref{eq:nbd_bispectrum}) is clearly different from the other two. While for features (eq. \eqref{eq:feat_bispectrum} and eq. \eqref{eq:res_bispectrum}) the argument in the oscillating functions explicitly depends on the sum all three comoving momenta, the argument in eq. \eqref{eq:nbd_bispectrum} depends on the ratio of momenta. Consequently we can expect a rather small overlap. This becomes even more apparent once we adapt a new set of variables
\begin{eqnarray}
k = k_t/2,&\:\:\:&k_1 = k(1-\beta) \nonumber \\
k_2=\frac{1}{2}k(1+\alpha+\beta)&\:\:\:&k_3=\frac{1}{2}k(1-\alpha+\beta)\nonumber \\ 
dk_1 dk_2 dk_3 &=& k^2dk d\alpha d\beta \nonumber,
\end{eqnarray}
proposed in \cite{Fergusson}. As a consequence the argument in eq. \eqref{eq:nbd_bispectrum} will depend on the two variables $\alpha$ and $\beta$, while the arguments in eq. \eqref{eq:feat_bispectrum} and \eqref{eq:res_bispectrum} will only depend on $k$. In that sense, we can say that oscillations in these shapes are in  {\it orthogonal} directions. 

\begin{figure}[t]
   \centering
   \includegraphics[scale=.56]{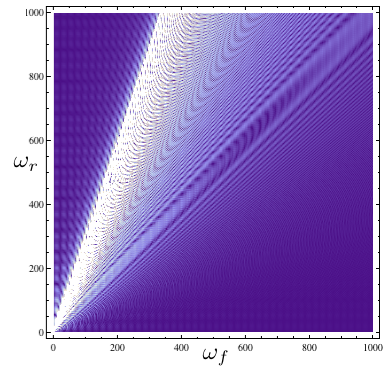} 
   \caption{The correlation between the bispectral shapes of \protect\eqref{eq:feat_bispectrum} and \protect\eqref{eq:res_bispectrum} for various values of the frequency. The light areas correspond to correlations of order $\mathcal{O}(1)$, while the dark shaded areas correspond to correlations close to 0. The correlation was computed with $\delta =\gamma_1=0$.} 
    \label{fig:cosine_feat_res}
\end{figure}

In addition, for both the feature and resonant bispectrum the frequency is fixed along one direction. That is, the frequency does not change (feature) or only slightly changes (resonant) when you run through a fixed direction in comoving momentum space. For the non-BD bispectrum however the argument in the oscillating function has a component that scales as $1/x_j$. Consequently for $x_j\rightarrow 0$ the effective frequency $\omega_{eff}\rightarrow \infty$. Naturally, $x_j$ is cutoff from below (as $k_{min}/k_{max}\sim10^{-4}$), however even with a cutoff the range in effective frequencies is large along a direction. This effect is present at all frequencies, and it turns out it will determine the efficiency of mode expansion for this bispectrum discussed in the next section. 

We have numerically calculated the correlator as defined in eq. \eqref{eq:correlator} between both feature bispectra and non-BD spectrum. We found the correlation to be maximal for low values of both frequencies (of order 1 percent around $\omega =10$), indicating that there is no evidence for a particular resonant frequency; the largest correlation occurs due to the fact that there are less oscillations, thereby decreasing the chance for (almost perfect) cancelations in the integral. As expected, we can safely conclude that these shapes are distinguishable/orthogonal. 

For the two bispectra of eq. \eqref{eq:feat_bispectrum} and \eqref{eq:res_bispectrum} we can expect a larger correlation. The appearance of a $\log$ in eq. \eqref{eq:res_bispectrum} is the only major difference between the two bispectra. In the new coordinate set, the bispectrum of \eqref{eq:feat_bispectrum} does not depend on the $\alpha$ or $\beta$.  Let us try to make a simple analytical approximation of the relevant correlator before we compute the correlation numerically. The first term in \eqref{eq:res_bispectrum} dominates the second for large values of $(f\phi_*)^{-1}$. Therefore for simplicity we neglect the second term. As a consequence both terms now depend only on $k$. In the computation of the correlator the integration over $\alpha$ and $\beta$ drops out and to get an indication of the resonance we only need to investigate the following integral:
\begin{eqnarray}
\int_0^3 x_t dx_t \sin ( \omega_f  x_t+\delta) \sin (\omega_r \log x_t+\gamma_1).
\end{eqnarray}
where we assumed that at most $k_t = 3k_{max}\rightarrow x_t  = 3$. This integral can be done analytically and results in a sum of $\Gamma$ functions (we have set $\delta=\gamma_1=0$). The interpretation of the result is rather complicated as all terms are divergent and there are no terms that can be easily neglected. However, one can plot the result and find that there is a  clear resonance `area' around $\omega_r \simeq 20\omega_f$. We have confirmed this resonance as a function of frequency when considering the full expression and allowing both phases to be non-zero.  We have plotted (fig. \ref{fig:cosine_feat_res}) the correlation for a range of frequencies ($10<\omega<1000$) and a phase $\delta =\gamma_1=0$. The largest values obtained from this numerical computation are of order 0.6, or 60 percent correlation (we have used discreet steps of $\delta \omega =10$), and we expect there to exist correlation of $\mathcal{O}(1)$  for some specific values of $\omega$). As such it will be hard to discriminate between these two models solely using observations of the bispectrum (as one could simply confuse frequencies). However, as mentioned before, axion inflation for example predict a large scalar to tensor ratio. Measurement of $r$ could break the degeneracy between a sharp feature in the potential versus axion inflation. In addition, one does not expect $\omega_r<10$ since it would not produce observational $f_{NL}^{res}$, while for the feature bispectrum the natural frequency is no larger than $\omega_f\sim 50$. If one would be able to extract a frequency from the data, a large frequency would favor a resonant model while a low frequency could indicate a sharp feature. 

\vspace{30mm}


\section{Mode Expansion}\label{Mode_expansion}

\subsection{Power Modes}\label{Powermodes}

The discussed primordial bispectra have very little in common with the constrained local, equilateral and orthogonal bispectra. Typically, to constrain any type of non-Gaussianity one computes the correlator  (eq. \eqref{eq:correlator}) and derive the so-called `fudge' factor which indication how much `signal' leaks into an existing template With the use of the fudge factor one is able to deduce a bound on the amplitude of the unconstrained bispectrum. The reason why certain templates have been constrained and some others have not, is two-fold. First and foremost, until now most models produced non-Gaussianities that can roughly be placed in one of the constrained types. For this reason, it was not immediate to search for any other type, simply because there were no models that indicated bispectra with completely orthogonal characteristics. Of course, optimally, one would simply look for the full bispectrum as a function of the multipole numbers instead of constraining the amplitude in particular bispectral configuration, but the low S/N and computational limitations have so-far restrained us to the former. 

The second reason not to look for more `exotic' bispectra is that for a fast estimator, the bispectrum one would like to constrain needs to be factorizable and scale invariant. That is, it is useful if the bispectrum can be we written as sum of products of functions, where each function only depends on one direction in multipole or comoving momentum space. It has been shown that such factorizability reduced the number of computations one has to make in order to constrain the amplitude of the bispectrum by a factor $l^2$, where $l$ is the number of observable multipoles of the experiment (leaving only $l^3$ computations). 

The constrained non-Gaussian amplitudes (in the form of $f^i_{NL}$, where $i$ labels the comoving momentum type, local, equilateral or orthogonal) are all based on templates that are factorized in the manner explained above. For instance, although DBI inflation does not produce a factorized bispectrum, it is well approximated by the equilateral template \cite{Creminelli:2005hu}, that is factorized by construction. The same is true for both the local and orthogonal template, as well as the enfolded \cite{Meerburg2009a} template. However, the method for constructing such factorized approximations of existing theoretical bispectra is rather ad-hoc. Until recently there was no procedure no construct a factorized bispectrum using a consistent prescription.

In \cite{Fergusson:2009nv}, a method for constructing factorized approximations to theoretical bispectra has been proposed using polynomial expansion. The approach is fairly straightforward; one defines a set of orthonormal 3 dimensional functions (where orthonormal is defined using a correlator of the form\footnote{For the construction of these polynomial modes we set $w=1$. Once computing the correlator between the original and the reconstructed spectrum one can take $w=1/k_t$ in order to see how much of an effect projection onto multipole space can have. We find that it reduces the correlation by $5$ to $10\%$ in both polynomial expansion and Fourier expansion. As such, it should not effect the conclusions we draw in this paper where all correlation shown are based on $w=1$. In order to build modes that are optimized for multipole expansion you should start by considering a weight function $1/k_t$. This is beyond the scope of this paper.}    eq. \eqref{eq:correlator}, and the weight function can be adjusted) which are a-priori factorized and from there one computes the corresponding weight factors ($\alpha_n$) via the inner product between a number of polynomial modes ($R_n$) up until a sufficient overlap between the polynomial expansion and  the original bispectrum is established, i.e. until $N$ such that 
\begin{eqnarray}
S(x_1,x_2,x_3)&\simeq& \sum_{n=0}^N \alpha_n R_n(x_1,x_2,x_3).
\end{eqnarray}
Without discussing the details of constructing such polynomial modes (see \cite{Fergusson:2009nv} for a detailed description), here we want to try and investigate how well this would work in case of oscillatory bispectra of  \eqref{eq:feat_bispectrum}, \eqref{eq:res_bispectrum} and \eqref{eq:nbd_bispectrum}.

Before we do so, let us make a few notes. First of all, recall that the objective of the expansion is to factorize a given theoretical bispectrum. However, as you can see from eq. \eqref{eq:feat_bispectrum}, this particular bispectrum, albeit a best-fit approximation\footnote{The proposed envelop function has the form $(k_1+k_2+k_3)^n e^{(k_1+k_2+k_3)/k_*m}$,, where $m$ and $n$ are fitted to the numerical results. The envelope function is therefore also factorizable. Again, we did not consider this envelop since it is smooth compared to the oscillatory part of the bispectrum. However, such an envelope could be of significant influence in predicting the correlation in multipole space \cite{Fergusson}.}, is already of the factorized form. One can still try to expand this in terms of power law polynomials, as described here, since polynomial modes will in general behave better numerically. The other two examples of primordial bispectra are not factorizable in terms of oscillating functions using simple identities. Consequently, the polynomial expansion seems to be a good first effort in order to set up an approximately factorized form. 

Secondly, were we able to expand these into a factorized form, and subsequently projected to multipole space and applied to the data, we might still miss the entire signal, simply because one of the free parameters is the frequency of the oscillations. For a non-BD bispectrum and the axion inflation model, the range of possible frequencies spans (at least) 2 orders of magnitude. Therefore, if we would fix the frequency, searching for a signal with a constructed factorized template would probably not be the best approach. Fortunately, we will later see that if you would measure mode functions in the data, instead of a fixed template, one could in principle extract information about a variety of oscillating signals. Let us emphasize that even if we would not be able to reconstruct a factorized form of a given spectrum with a small number of modes, it is still very well possible we could observe the same spectra by measuring a small number of mode functions in the data (effectively the frequency (and the phase) remain a free parameter during mode extraction). 

\subsubsection{Feature Bispectrum}\label{feature_spectrum}

\begin{figure}[b]
   \centering
   \includegraphics[scale=.65]{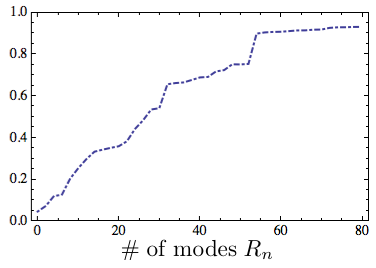} 
   \caption{Example of the increasing correlation (eq. \protect\eqref{eq:correlator} between the approximation of eq.  \protect\eqref{eq:feat_bispectrum} and the original spectrum. Here $\omega_f=9$ and we find that it requires over 80 modes to achieve perfect correlation.}
    \label{fig:power_correlation}
\end{figure}

First we consider the bispectrum coming from a feature in the potential (eq. \eqref{eq:feat_bispectrum}). Out of the given examples it has the simplest form (excluding the envelope). We choose $\delta =0$ for simplicity, and since the phase can always be scaled out it will not affect the results\footnote{One would also have to consider $\cos \omega_f x_t$ but we found no difference when expanding between the cosine and sine in terms of the required number of modes.}. In table \ref{tab:number_of_modes} we have computed the number of modes necessary to get a correlation of  at least $98\%$ with the original spectrum for several values of $\omega_f$. As expected, as the frequency is increased, one has to expand the bispectrum with a (rapidly) growing number of modes. For $\omega_{f} =9$ we get a $93\%$ correlation with 82 modes. On itself, it actually quite remarkable that one is able to reproduce the spectrum with a limited number of modes. Recall that the possible feature at $l\sim30$ would result in a (decaying) oscillation with $\omega_f\sim50$, which would be hard to fit this way. On the other hand, as we argued earlier, a frequency of $\omega_f=50$ can be considered an upper limit, as features at higher multipole number would result in longer wavelengths. We have plotted an example of how the correlation between the original spectrum and the expansion increases as a function of the number of mode functions in the expansion in Fig. \ref{fig:power_correlation}.
\begin{table}
\centering
\begin{tabular}{ |c|c|c|c|c|c|c|c|c|c|} 
 \hline\hline	
 $\omega_f$ & 1 & 2 & 3& 4& 5& 6& 7&8&$9^*$\tabularnewline	
 \hline
 $\#$ of modes $R_n$&1&5& 8&12&18&43&55&69&82\tabularnewline
 \hline 
 \end{tabular} 
 \caption{ As the frequency is increased it requires a rapidly growing number of modes to get over $98\%$ correlation with the original spectrum.}
 \label{tab:number_of_modes} 
 \end{table}

\subsubsection{Resonant Bispectrum}\label{resonant_bispectrum}

\begin{figure}
   \centering
   \includegraphics[scale=.65]{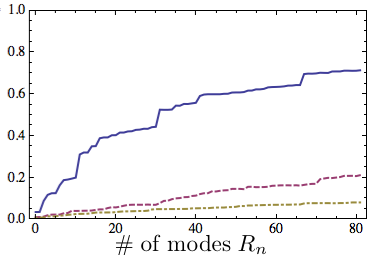} 
   \caption{The correlation between $\sin ( \omega_r \ln x_t)$ for 3 different frequencies. From top to bottom $\omega_r=20,40$ and $60$. Beyond frequencies of $60$ polynomial expansion would require many modes to achieve significant correlation with the original spectrum. }
    \label{fig:axion_polynomial}
\end{figure}

Next, let us consider the resonant bispectrum. It is quite similar to the feature bispectrum, but theoretically we expect much larger frequencies ($20\leq \omega_r\leq 10^3$). We have computed (figure \ref{fig:axion_polynomial}) the correlation between the expansion and the original spectrum, chosen to be $\sin (\omega_r \ln x_t)$ since for the same reason as before the phase will barely affect the number of modes required to reconstruct the spectrum. As expected, the convergence of the correlation towards one (perfect overlap) proceeds slowly. For the lowest frequency we considered ($\omega_r =20$), the correlation reaches $71\%$ after 82 modes. For $\omega_r=60$ the largest correlation we can achieve is $7\%$ after 82 modes. Recall that  the amplitude of the resonant bispectrum is proportional to its frequency. The maximum correlation between existing templates and the axion spectrum is of order $1\%$ \cite{Flauger:2010ja} (although for small frequencies this can be $10\%$ for the equilateral template) and possibly measuring these modes in the data would therefore still allow for a constraint on axion inflation that is 10-100 times\footnote{This would require adding the modes once extracted from the data.} better than what we have now. In general, increasing the frequency above $\omega_r\sim60$, large correlation becomes hard to achieve with a limited number of modes.

\begin{figure}
   \centering
   \includegraphics[scale=.65]{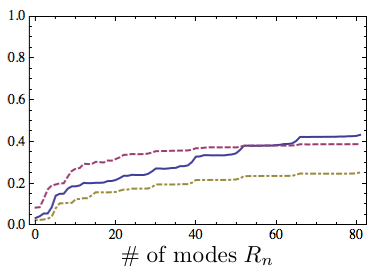} 
   \caption{The correlation between the non-BD bispectrum of eq. \protect\eqref{eq:nbd_bispectrum} and its polynomial expansion as a function of mode number for 3 different frequencies $\omega_v=20$ (solid), $40$ (dashed) and $60$ (dot-dashed). We have set $\gamma_2=0$ but have found very little difference for non-zero $\gamma_2$. }
    \label{fig:non_bd_pol_modes}
\end{figure}

\subsubsection{Non-BD Bispectrum}\label{nonBD_bispectrum}

For the non-BD spectrum of eq. \eqref{eq:nbd_bispectrum} we find that computing the correlator numerically requires a very high resolution, because this shape contains terms that are singular and the spectrum as a whole is only finite due to the exact cancellation between all the specific terms. To avoid these problems one has to stay away from the line(s) $x_{j+1}+x_{j+2}=x_j$, which can be done by adding a small $\epsilon$ in the vicinity of this line in the integral that defines the dot product (eq. \eqref{eq:dotproduct}). The results are shown in figure \ref{fig:non_bd_pol_modes}. Even for low frequency ($\omega_v=20$) we can not achieve a large correlation with $82$ modes. On the other hand, increasing the frequency does not really affect the ability to reach similar correlation. Overall, we find that the non-BD bispectrum is the most difficult to reconstruct due to the appearance of terms that diverge inside the argument, since for $x_j\rightarrow 0$ the frequency of the signal becomes extremely large at some of the edges of the tetrahedral domain. The observation that we can still reach some correlation is because there are also areas on the tetrahedral domain where the effective frequency is relatively small. These areas remain even if $\omega_v$ increases (al though they should become smaller and smaller) Consequently we find that achievable correlation with 82 modes is small but does not decrease significantly when you go to higher frequencies. The correlation with smooth spectra is typically of order $\ll1\%$ \cite{Meerburg2009b} which means that an accumulated correlation of only a few percent could drastically improve the constraints we can put on $|\beta|$ as the amplitude scales proportionally to $\omega_v^3$. 

Polynomial expansion seems to work reasonably well for low frequencies of the various bispectra. For larger frequencies, to reconstruct the original spectrum the polynomial expansion requires an increasing number of modes. Given the large allowable frequencies for the resonant and non-BD bispectra, polynomial expansion might not be the most effective way of expanding. In the next section we will explore another type of expansion which uses a Fourier basis.  We will investigate if such a basis would require less modes to achieve similar correlation.

\begin{table*}
\centering
\begin{tabular}{|c|c|c|c|c|c|c|}
\hline\hline 
$n=0\rightarrow000$ & $n=4\rightarrow111$ & $n=8\rightarrow022$ & $n=12\rightarrow113$ & $n=16\rightarrow222$ & $n=20\rightarrow024$ & $n=24\rightarrow133$\tabularnewline
\hline 
$n=1\rightarrow001$ & $n=5\rightarrow012$ & $n=9\rightarrow013$ & $n=13\rightarrow023$ & $n=17\rightarrow123$ & $n=21\rightarrow015$ & $n=25\rightarrow124$\tabularnewline
\hline 
\multicolumn{1}{|c|}{$n=2\rightarrow011$} & $n=6\rightarrow003$ & $n=10\rightarrow004$ & $n=14\rightarrow014$ & $n=18\rightarrow033$ & $n=22\rightarrow006$ & $n=26\rightarrow034$\tabularnewline
\hline
$n=3\rightarrow002$ & $n=7\rightarrow112$ & $n=11\rightarrow112$ & $n=15\rightarrow005$ & $n=19\rightarrow114$ & $n=23\rightarrow223$ & $n=27\rightarrow115$\tabularnewline
\hline
\end{tabular}

\caption{The association of mode numbers for the first 27 modes. For example $n=10$ corresponds to the mode for which one direction is of maximally 4th order and the other two are $0$ order, i.e. for polynomial modes $R_{10}\propto f(1,x_1,x_1^2,x_1^3, x_1^4)+f(1,x_2,x_2^2,x_2^3, x_2^4)+f(1,x_3,x_3^2,x_3^3, x_3^4)$.}
\label{tab:mode_numbers}
\centering
\end{table*}

\subsection{Fourier Modes}\label{Fouriermodes}

The polynomial expansion of \cite{Fergusson:2009nv} is  based on power modes, i.e. the expansion is in increasing order of $x^n$. This is not necessarily optimal for describing oscillatory functions. There are two possible alternatives; the first one would be to expand the argument into a sum of functions, that each depend on one direction only. A such, one can again use trigonometric identities to expand the cosine and sine into factorized forms (be that oscillatory functions). The second option could be to use a  Fourier expansion instead of a polynomial expansion. This would only be useful if for large frequencies you would need a small(er) number of modes. Before we get into Fourier mode expansion let us briefly discuss the alternative of expanding the argument in the oscillatory function. 

This option would only suffice if the approximation requires 2 modes maximally. If it requires more modes, you will get product of two or three different directions in momentum space, and as a result you will not be able to expand the cosine and the sine. Let us consider the axion model. The argument is given by $\omega\log x_t$. Using the mode expansion, one finds that one can achieve $>99\%$ correlation after just two polynomial modes; zero order and first order. Not surprisingly this is almost equivalent to a Taylor expansion to first order of  $\log k_t/k_*$ around the point $k_t\sim 1.4\sim\sqrt {2}$. Consequently, there are no cross-terms, and one can expand the cosine and sine into factorizable function of the three comoving momenta, just like you could expand the feature spectrum into oscillating functions. As it turns out however, although there is a $99\%$ correlation between the arguments after expansion, the full bispectrum is very sensitive to small deviations in the argument, especially for large frequency. Consequently, the correlation between the full bispectrum and the approximated bispectrum decreases as a function of the frequency; from $\sim90\%$ for $\omega_r =1$ to $\sim 50\%$  for $\omega_r=20$. Although this is equivalent to what can be achieved with the polynomial expansion using $~7$ modes, the problem is that we can not improve it in any way. Since this will only work for a first order expansion, we can never reach beyond  $50\%$ correlation, unlike the polynomial expansion, where we can simply include more modes.  Note that for non-BD model this method will not work as the argument is already a product of two directions in comoving momentum space, i.e. $(k_{j+1}+k_{j+2})/k_j$.  

\begin{figure}
   \centering
   \includegraphics[scale=.57]{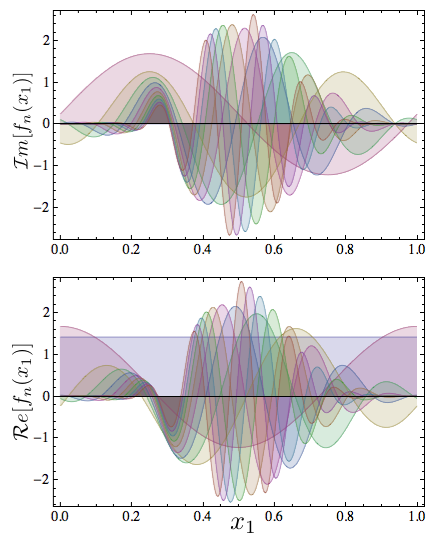} 
   \caption{The one dimensional orthonormal Fourier functions $f_n(x_1)$ within the tetrahedral domain for the first 11 modes. }
    \label{fig:Qmodes}
\end{figure}

The second option is to consider a Fourier expansion, where we try and fit terms such as $\sin \omega f(x,y,z)$ to a sum of Fourier modes that all depend on one direction only. Such factorization would still lead to the $l^2$ reduction in computation, since the integrals in $k$ space can now be performed individually\footnote{In \cite{Fergusson:2009nv} Fourier mode expansion is briefly discussed in section E as a possible orthonormal basis, however no results are shown.}. We consider $\exp [i 2  \pi n x]$ as our basis function (as apposed to $x^n$) and constructed a orthogonal set of three dimensional mode function similar to \cite{Fergusson:2009nv}. The first few one dimensional functions are given by
\begin{eqnarray*}
f_0(x_1) &=& \sqrt {2}\\
f_1(x_1) &=& 0.22+0.23 i + 1.45 e^{2 i \pi x_1}\\
f_2(x_1) &=& -0.0087+0.041 i (0.088+0.62 i)  e^{ 2 i \pi x_1}- \\
&&(0.31+1.12 i) e^{ 4 i \pi x_1}\\
f_3(x_1) &=&(-4.9+2.2i)10^{-3} -(0.15-0.11i) e^{ 2 i \pi x_1}+\\
&&(0.68-0.65i)  e^{ 4 i \pi x_1}-(0.59-0.65 i) e^{6 i \pi x_1}\\
f_4(x_1) &=& (-6.5-3.8i)10^{-4}-(0.042+0.017i) e^{2 i \pi x_1}+\\
&&(0.44+0.12i)  e^{4 i \pi x_1}-(1.+0.2i)e^{6 i \pi x_1}+\\
&& (0.63+0.1i)e^{ 8 i \pi x_1}\\
f_5(x_1) &=& (-1.1-11.1 i)10^{-5}-(0.002+0.011i)e^{ 2 i \pi x_1}+\\
&&(0.051+0.16 i)e^{4 i \pi x_1}-(0.25+0.63 i)e^{6 i \pi x_1}+\\
&&(0.4+0.89 i)e^{ 8 i \pi x_1}-(0.21+0.41i)e^{ 10 i \pi x_1}\\
...,
\end{eqnarray*}
The functions are shown up to $n=10$ in figure \ref{fig:Qmodes}.  From these one can construct the three dimensional basis functions via a product of each mode and symmetrization of three comoving momentum arguments; $x_1$, $x_2$ and $x_3$
\begin{eqnarray}
\mathcal{Z}_{prs}(x_1,x_2,x_3)&\propto&[f_p(x_1)f_r(x_2)f_s(x_3)+\mathrm{5\;perm}].\:\;\;\;\:
\end{eqnarray}
One has to introduce a counting scheme to re-numerate the three labels $\{p,r,s\}$ to $n$. We have chosen equal slicing counting \cite{Fergusson:2010dm}, of which the first $27$ modes ($n$) and their association ($\{p,r,s\}$) are shown in table \ref{tab:mode_numbers}.  After the construction of these modes, one has to apply additional Gramm Schmidt orthogonalization to $\mathcal{Z}_n$ to increase orthonormality of different mode functions.  We refer to the three dimensional orthonormalized modes as $\mathcal{F}_n$ and the corresponding weights as $\tilde{\alpha}_n$. 
\begin{eqnarray}
S(x_1,x_2,x_3)&\simeq& \sum_{n=0}^N \mathcal{R}e\left(\tilde{\alpha}_n \mathcal{F}_n(x_1,x_2,x_3)\right).
\end{eqnarray}
If $S$ would have been complex, one should add $ i \mathcal{I}m\left(\tilde{\alpha}_n \mathcal{F}_n\right)
$ in order to take this into account. The coefficients $\tilde{\alpha}_n$ can be computed by taking the inner product (eq. \eqref{eq:dotproduct}) between the original shape function (bispectrum) and the various mode functions $\mathcal{F}_n$, i.e.
\begin{eqnarray}
\tilde{\alpha}_n &=&  \int_{\Delta_{x_i}} dx_1 dx_2 dx_3 S(x_1,x_2,x_3) \mathcal{F}^*_n (x_1,x_2,x_3).\;\;\;\;\;
\end{eqnarray}

\vspace{20mm}

\begin{figure}[b]
   \centering
   \includegraphics[scale=.65]{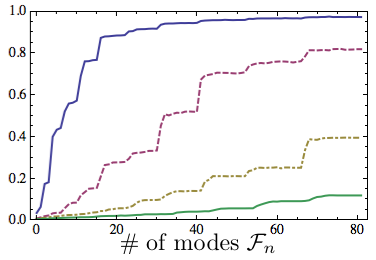} 
   \caption{Correlation between $\sin(\omega_r \ln x_t)$ for frequencies $\omega_r=20,40,60$ and $80$. Compared with the polynomial mode expansion we reach similar correlation using about 5 times less modes. Also note that the increase of correlation is somewhat discreet, indicating that we might need only a fraction of these modes to reconstruct the original spectrum. We will discuss this observation in the next section. } 
    \label{fig:axion_fourier}
\end{figure}
\begin{figure}
   \centering
   \includegraphics[scale=.55]{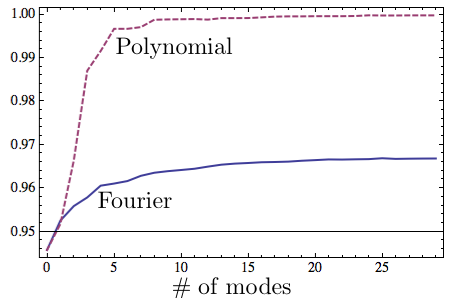} 
   \caption{The correlation between the DBI bispectrum and both polynomial and Fourier expansion as a function of the maximum number of modes. As expected, both expansions start out equally (the zero mode of the real part of the Fourier expansion is equivalent to the zero mode of the power law expansion), but while the power law reaches a correlation of $>99\%$ after just 5 modes, the Fourier remains stuck at $~97\%$.}
    \label{fig:fourier_vs_power}
\end{figure}
\begin{figure*}[t]
   \centering
   \includegraphics[scale=.65]{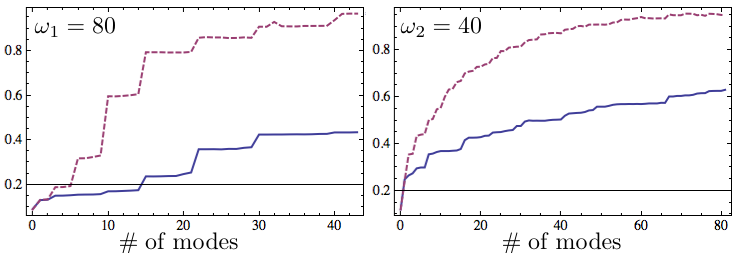} 
   \caption{The correlation as a function of mode number for two out of three toy-spectra $F_1$(left) and $F_2$(right) in eq. \protect\eqref{eq:toy_model}. In both cases Fourier expansion (dashed) leads to faster convergence compared to polynomial expansion (solid). }
    \label{fig:correlation_toy_spectra}
\end{figure*}

\subsubsection{Feature Bispectrum}\label{Feature_bispectrum_2}

For the feature bispectrum we do not necessarily have to consider the Fourier expansion\footnote{We will later show that we would also find it when we would search for resonant bispectra in the data, as the weights peak at almost the same mode numbers.} since that spectrum can be rewritten into a product of Fourier modes simply by using trigonometric identities, e.g. 
\begin{eqnarray}
\sin \omega_f x_t &=& \cos \omega_f x_3\left(\sin \omega_f x_1\cos \omega_f x_2+\right.\nonumber\\
&&\left.\cos \omega_f x_1 \sin \omega_f x_2\right)+\nonumber\\
&& \sin \omega_f x_3 \left(\cos \omega_f x_1 \cos \omega_f x_2\right.-\nonumber\\
&&\left.\sin\omega_f x_1 \sin \omega_f x_2\right.
\end{eqnarray}

The other two bispectra are not of the same form, since their arguments are non-linear functions, i.e. $\ln x_t$ for resonant non-Gaussianities and $(x_{j+1}+x_{j+2})/x_j$ for for non-BD modifications and these can be made of the form above by expanding, using the constructed Fourier modes $\mathcal{F}_n$. Given the form of the first argument you expect only a limited number of modes to significantly contribute, for example those modes that have equal mode number in the directions $x_1$, $x_2$ and $x_3$ (you should think about this expansion as a series around the point $x_t$, see table \ref{tab:mode_numbers}). For the second argument you expect more modes to matter, since the arguments depend on all three directions independently. Consequently the weights $\tilde{\alpha}_n$ are expected to be close to zero for many $n$ when expanding resonant non-Gaussianities, while for the non-BD scenario they should all matter to some extend (and obviously more modes will be important for large $\omega_v$). 

\subsubsection{Resonant Bispectrum}\label{resonant_bispectrum_2}

We have computed the correlation for the axion bispectrum with the Fourier expansion for frequency ranges of $\omega_r=20-80$ up to 82 modes (figure \ref{fig:axion_fourier}). As expected, we see that there are only a few modes that give significant contribution to the correlation, while most modes give only very little contribution and are not important for the expansion. We will discuss this fact in the context of CMB data mode extraction in the section 4. Given that the allowed range of frequencies $20\lesssim \omega_r \lesssim 10^3$ this expansion is actually reasonable for the lower frequencies and the number of modes necessary to establish similar correlation as the polynomial expansion is reduced by a factor 5.

\subsubsection{Non-BD Bispectrum}\label{nonBD_bispectrum_2}

As for the polynomial basis expansion, the presence of a large number of features in the non-BD bispectrum does not allow for a fast reconstruction of the spectrum. In fact, expansion in the Fourier basis requires even more modes compared to the polynomial basis, reaching only $\sim20\%$ correlation after 82 modes with $\omega_v=20$. We also find that $\omega_f=40$ actually reaches a slightly larger correlation, although this seems mostly due to a relatively large correlation with the zero order ($n=0$) mode. Most likely this is caused by the fastest oscillating part of the spectrum which, in combination with numerics, could add constant power. We did observe something similar in figure \ref{fig:non_bd_pol_modes} for polynomial modes where the zero mode causes the correlation of the non-BD bispectrum reconstruction with $\omega_v=40$ to be better initially compared to bispectrum expansion with $\omega_f=20$. 

In most realistic scenarios $\omega_v>100$ (otherwise your effective field theory approach breaks down) and therefore both polynomial expansion and Fourier expansion fail to reconstruct this bispectrum effectively. The possible explanation why Fourier expansion is even worse than polynomial expansion for this type of bispectrum, seems to be related to the rapid change in frequency in a fixed direction. Fourier expansion is optimized for scale invariant frequencies. The polynomial expansion is simply optimized in reproducing as many different shapes as possible, explaining the observation that it is able to slowly increase correlation with the addition of modes while Fourier expansion seems to converge. Given the large enhancement of the amplitude $f_{NL}^{nBD}$ (which scales as $\omega_v^3$), one might still be able to extract some information from that data even with such small correlations.  

Another possibility is that once non-BD bispectrum is projected onto multipole space one might establish a larger correlation with fewer (multipole) modes. The projection has the tendency to wash out small features (hence the weight of $1/k_t$ in the correlator.). We hope to report on this in the future.

\subsubsection{Toy Spectra}

To investigate the power of the Fourier expansion for oscillatory bispectra we have also tried to fit three toy-model shapes moving in different direction through comoving momentum space
\begin{eqnarray}
F_1&=&\frac{1}{k_1^2k_2^2 k_3^2}\left(\sin \frac{\omega_1}{x_1+1}+\sin \frac{\omega_1}{x_2+1}+\sin\frac{\omega_1}{x_3+1}\right),\nonumber\\
F_2&=&\frac{1}{k_1^2k_2^2 k_3^2} \sin \omega_2 x_1 x_2 x_3,\nonumber\\
F_3&=&\frac{1}{k_1^2k_2^2 k_3^2}\left(\sin \frac{\omega_3 x_t}{x_1+1}+\sin \frac{\omega_3 x_t}{x_2+1}+\sin\frac{\omega_3x_t}{x_3+1}\right).\nonumber\\
\label{eq:toy_model}
\end{eqnarray}
We find again that for such a shapes the correlation increases about 5 times faster compared to polynomial mode expansion with the same frequency. The correlation as a function of mode numbers for $F_1$ and $F_2$ are shown figure \ref{fig:correlation_toy_spectra}. We will discuss the weights of these models in the next section. Note that $F_1$ is already of the factorized form, however here we simply aim at showing the effectiveness of Fourier expansion. We want to emphasize that these spectra are not based on any physical model, but simply show that in general oscillatory spectra are better fitted using a Fourier basis. 

\subsubsection{Smooth Spectra}

Although the Fourier expansion seems to work well for resonant non-Gaussianities and the toy-spectra, compared to polynomial expansion we confirm that Fourier expansion is not as effective: it is easier to gain fast convergence with a limited number of modes for most oscillating bispectra, but it is difficult to get correlation beyond $~0.97$ for smooth bispectra. This is probably due to overshooting at the boundaries as discussed in \cite{Fergusson:2009nv}. We explicitly show this in figure \ref{fig:fourier_vs_power} where we compare expansion of the `smooth' DBI inflation bispectrum (which is very similar to equilateral), using Fourier modes and polynomial modes.

We conclude that Fourier expansion is a viable alternative for polynomial expansion in the case of oscillatory bispectra with relatively large frequencies. Using the Fourier expansion we can achieve factorizabilty of various oscillating bispectra with significantly less modes compared to polynomial expansion. For frequencies $\omega\gg 50$ polynomial and Fourier expansion are both unable to reconstruct the original spectrum with a small number of modes. In order to reconstruct models with such large frequencies, one should look for alternative methods. However, constraining these models with only limited number of modes seems to be a practical possibility. This will be topic of the next section.

\begin{figure}[b]
   \centering
   \includegraphics[scale=0.55]{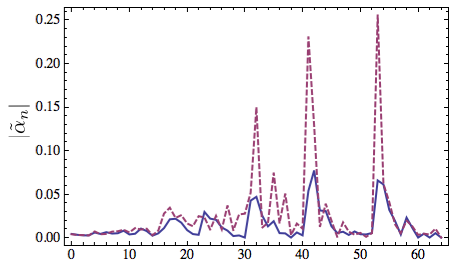}
   \caption{The weights derived for 65 modes for both $cos \omega_r  \ln k_t$ (dashed)  with $\omega_r=50$ and $\sin \omega_f k_t$ (solid) with $\omega_f= 20$ showing that these both peak for similar mode numbers. Although distinguishing between these would be quite hard, it seems that for the feature bispectrum the values of the weights $\tilde{\alpha}_n$ are peaked sharper.}
    \label{fig:axion_vs_feature}
\end{figure}

\begin{figure*}
\centering
\includegraphics[scale=0.60]{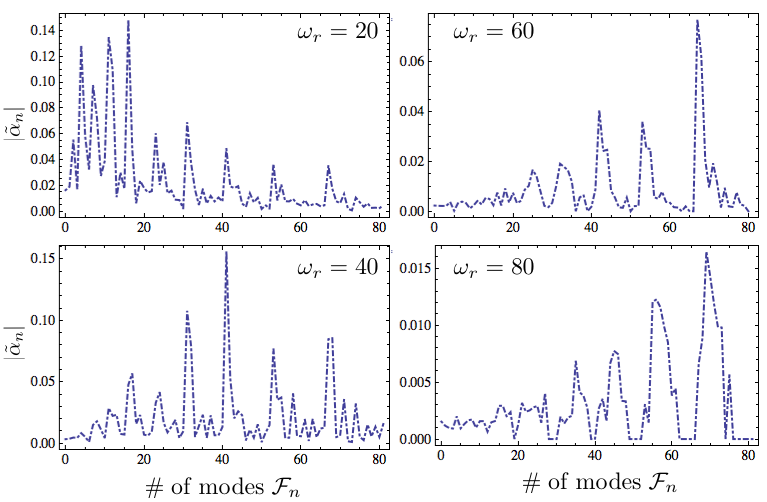}
   \caption{The weights $\sqrt{\alpha_n \alpha_n^*}$ for a resonant bispectrum as a function of mode number for various frequencies. It is clear that only limited number of modes are valuable in the reconstruction of the original spectrum via mode expansion. From an observational point of view this is very convenient as it would require the measurement of only a limited number of Fourier modes to learn about oscillations in the primordial bispectrum.}
    \label{fig:alphas}
    \end{figure*}

\section{Discussion}\label{discussion}

Even though the expansion of the oscillatory primordial bispectra becomes unavailing for really large frequencies, there are a number of interesting observations which could make constraining and expanding oscillating bispectra much more viable than presently argued. First of all, as predicted, the expansion in mode functions of the resonant bispectrum has a very discrete character; basically if you consider fig. \ref{fig:axion_fourier} only few modes actually contribute significantly to the convergence of the correlation. In fig. \ref{fig:alphas} we show the various weights ($|\tilde{\alpha}_n|$) as a function of mode number (as well as for $F_1$ and $F_3$ (not shown)). We can trace back the corresponding mode numbers in table \ref{tab:mode_numbers}. For instance there is a clear peak at $n=16$, which correspond to all directions being maximally of quadratic order, and $n=41$ with all directions being maximally of cubic order. Other peaks (e.g. $n=21$, $32$ and $53$) correspond to the modes in which two out of three directions have one and two maximal orders less than the third, i.e. in mode number $n=23$ two directions are maximally quadratic and the third is maximally cubic. As we already argued the location of these peaks makes sense, since the resonant model is a function of $k_t$ (or $x_t$), which is the sum of the three comoving momenta. Effectively this shape is orientated in the $k_t$ direction. One could only try to expand the spectrum only in those modes, which could significantly reduce the number of modes necessary. Since the important modes seem to be related to the direction of propagation of the oscillation, we find that this conclusion is independent of the phase. In other words, only the value of the weights will differ, not the mode numbers that are relevant for the expansion. This can be explained as follows. Consider a very simple example of an oscillating mode $\mathcal{R}e[e^{i (x+\delta)}]$. If we would expand this into polynomial mode functions, $\{1,x,x^2,...,x^n\}$ we would find that $\alpha_n$ would change as a function of $n$ if we vary $\delta$. This makes perfect sense, since we know the polynomial expansion of these functions exactly, as they are the Taylor series of the sine and cosine. If we would expand in Fourier modes $\{1,e^{ix},...e^{i n x}\}$, the expansion is obviously much simpler. However, more importantly, the complex phase will not affect the quantity $\alpha \alpha*$. Let us consider the mode with the largest value $\alpha \alpha*$ the resonance peak. The location of this resonance peak will be unaltered by a change of phase. For the weights of a polynomial mode expansion this is not true, as the introduction of a non-zero phase will cause this example to shift from a cosine to a sine, thereby transferring power from odd to even modes. This would cause peaks in $\alpha_n$ to shift from $n$ to $n+1$.

Secondly, from an observational point of view, given the discreteness of the correlation it is (obviously) not necessary to constrain all mode functions in the CMB data to get an indication of there is an oscillatory three point signal and what the possible frequency of this signal might be.  For resonant non-Gaussianities we only need to consider those modes that have a significant weight $\tilde{\alpha}$, and the measured value of the weights would be a direct measure of the frequency. If one could extract the multipole projected Fourier modes that are responsible for most of the weight, this could in principle provide signatures of primordial bispectra with frequencies much larger than $\omega_r=80$. Measuring modes up to e.g. $n=100$ would not only provide information about the frequency of the signal, but could also hint on the type of primordial bispectrum. The distinction between the feature bispectrum \eqref{eq:feat_bispectrum} and the resonant bispectrum \eqref{eq:res_bispectrum} would be more difficult, since the values of the weights peak at similar mode numbers although we have found that expanding the feature bispectrum in the constructed Fourier basis (instead of the simple trigonometric expansion discussed in section \ref{Feature_bispectrum_2}) could still be used to discriminate between the two signals (see figure \ref{fig:axion_vs_feature}). 
    \begin{figure*}
 \includegraphics[scale=0.60]{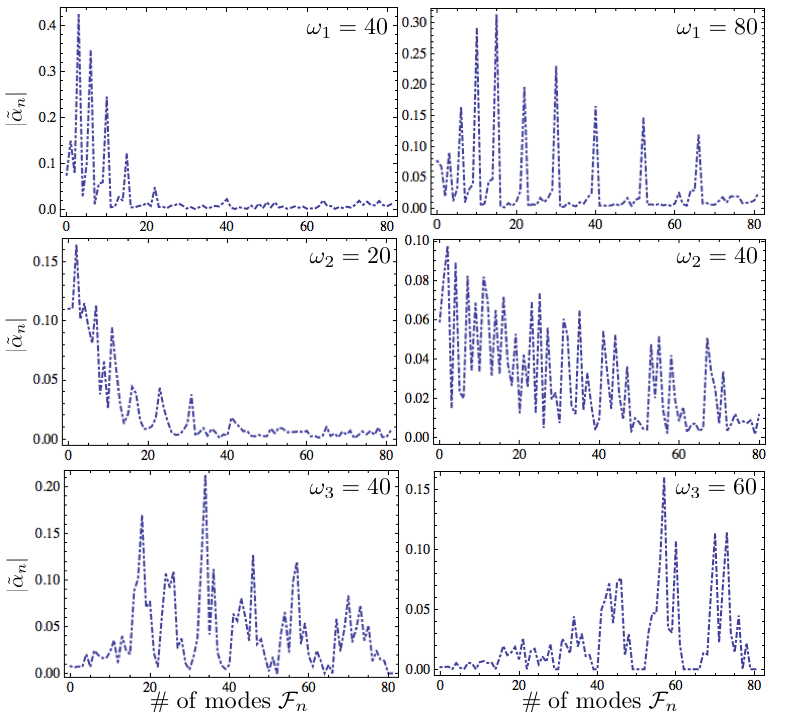}
   \caption{The weights $\sqrt{\alpha_n \alpha_n^*}$ for the three toy-spectra as a function of mode number for various frequencies. For the top spectrum the most relevant modes are those that have a maximal frequency of $\sim2\pi n$ for one comoving vector and are constant for the other two. For the second toy-spectrum (middle) the symmetry of the argument makes all weights relevant (decreasing as a function of mode number). The third example (bottom) has modes dominating of which two comoving momentum vectors have non-zero (equal) frequency and the third one is constant. For example, the lower-right bottom example has $|\tilde{\alpha}_{57}|$ dominating, which corresponds to $\{p,r,s\}=\{0,5,5\}$.}
    \label{fig:alphas2}
\end{figure*}

To emphasize the ability to extract information on the primordial shape solely from the modes that are important, we have investigated three toy-model shapes of eq. \eqref{eq:toy_model}.
We have computed the Fourier weights for two different frequencies in figure \ref{fig:alphas2}. As expected, $F_1$ has weights that peak when only one comoving momentum in in $k$ space is non-zero is, i.e. it peaks at the modes where one momentum oscillates and the other two momenta are constant (see \ref{tab:mode_numbers}). The obvious reason is that each term in $F_1$ depends on one comoving momentum variable only, implying that there should be no cross terms in the expansion. For $F_2$ we find that many more modes are relevant, which makes perfect sense given that the argument in the sine depends on all  three vectors in comoving momentum space. For $F_3$ however the argument effectively only depends on two comoving momenta, therefore the relevant mode functions (the ones with the largest $|\tilde{\alpha}|$) are the ones that have similar frequency in two momentum vectors and are constant in the third. 

In this paper we have only discussed mode functions in momentum space, and one either has to construct similar Fourier modes in multipole space or project these modes forward using the transfer function \cite{Fergusson:2009nv}, and use these to expand a late-time oscillatory bispectrum, and see if we get similar results in terms of mode number sensitivity. One expects that after projection the transfer function has caused some smoothing of the signal, which could render a Fourier basis less effective. On the other hand, intuitively it seems  perfectly reasonable that a Fourier basis should be much more efficient in reconstructing oscillatory bispectra from the data. In addition, the effects of the transfer function on the correlation in $l$ space can be examined by choosing the $1/k_t$ weight $w_k$ in the primordial correlation function. We have found that our results were only marginally affected when including this weight factor and therefore we expect that Fourier mode expansion should be equally efficient in multipole space. To make sure that this is actually true, we should compute the projection of several oscillatory bispectra and construct a orthonormal Fourier basis in multipole space. We hope to report on this in the near future. 

\section{Conclusions}\label{conclusion}

We have investigated the viability of mode expansion for bispectra that contain oscillations. The motivation for investigating such features and their mode expansion, is that recently it has been shown that several scenarios or mechanisms can produce such features not only in the power spectrum, but also in the bispectrum. The appearance of oscillations in the bispectrum makes comparison with existing bispectral constraints, based on smooth bispectra, very inefficient and there exists substantial room for improvement. In order to constrain oscillatory bispectra from the data, a logical first step is to factorize the bispectrum in order to efficiently compute its multipole counterpart. Polynomial expansion has been proposed to achieve factorization of a given theoretical bispectrum and we have investigated this for three different models. As expected, the larger the frequency of the primordial bispectrum, the more modes it requires to establish a reasonable approximation of the original spectrum. In the case of a feature in the primordial potential polynomial mode expansion might still be useful, at least for features at high multipoles (resulting in rather small frequencies in comoving momentum space). In fact, during the finalization of this paper the authors of \cite{Fergusson:2010dm} have considered a feature bispectrum and extracted 31 polynomial modes in the data, which allowed them to investigate late time bispectra with a maximal frequency of $\omega_f=5-10$ (in comoving momentum space). They did not find $3\sigma$ evidence for non-zero non-Gaussianity. The other two example bispectra typically have a lot more oscillations within the tetrahedral domain, resulting in many modes necessary to realize an acceptable correlation. Fortunately, both the resonant and non-BD bispectrum have an amplitude that scales with the frequency. Therefore, a small improvement in correlation could lead to a significant improvement in the ability to constrain the model by measuring these modes in the data and reconstructing the primordial signal.

Complementarily, we have proposed a different basis expansion, based on Fourier functions instead of polynomials.  This still leads to the necessary computational reduction one is after and therefore is a perfectly valid alternative. Such expansion is more relevant for resonant and non-BD scenario, since the feature bispectrum can already be transformed into Fourier modes analytically, using identities. We have shown that Fourier modes are much more efficient for the resonant bispectrum, reducing the number of modes necessary to establish the same correlation as polynomial modes by at least a factor of 5.  For the non-BD bispectrum both Fourier expansion and polynomial expansion are difficult. Correlation increases fast with the addition of modes, but quickly converges to a fixed value, where the fixed value decreases a function of frequency. We believe that this is due to the exact form of the bispectrum, which has many small features near the edges of the tetrahedral domain. One might hope that some of these very small features are washed out when you compute the multipole equivalent, although that would be very time consuming since the non-BD shape is not of the factorized form. We hope to investigate this in a future attempt.  In addition we have investigated three toy-spectra, not based on any particular model, which have a different oscillating orientation compared to the three theoretical models. Expanding these in Fourier modes show similar improvement compared to polynomial expansion as the resonant bispectrum. In general, we therefore belief that Fourier expansion is much more effective in the expansion of oscillatory spectra compared to polynomial basis expansion.

From an observational stand point, it seems that for resonant inflation only a limited number of modes contribute significantly in reproducing the original bispectrum. This allows us to consider only those modes that contribute substantially. This holds independent of the phase and frequency of the signal and is due to the specific form of this bispectrum, which oscillates (primarily) in the $k_t$ direction. Because the modes that are important for the reconstruction of the original bispectrum are independent of the frequency, this also implies that when one would observe these modes in the data one could in fact find evidence for much larger frequencies than discussed here, simply because for larger frequencies these modes will also matter but their respective weight will be smaller. Despite the fact that we could not optimally expand the non-BD bispectrum using Fourier modes, we did look into the three toy-sepctra. We found that other modes are important. Moreover, the modes that are important directly represent the orientation of the oscillating spectrum and could therefore discriminate between different bispectra quite effectively.  If this conclusion holds after forward projection into multipole space, measuring a number of Fourier mode functions in the CMB data would present an efficient way of deducing whether oscillations are present in the data and could give both an indication of the frequency and the shape of the primordial bispectrum. 

\medskip

\acknowledgments
The author would like to thank Jan Pieter van der Schaar, Pier Stefano Corasaniti, Ralph Wijers, Licia Verde, Ben Wandelt, James Fergusson, Xingang Chen and Michele Luguori for very helpful discussions and comments on the manuscript. He would also like to thank the hospitality of DAMTP, Cambridge, where this paper was finalized. The author was supported by the Netherlands Organization for Scientific Research (NWO), NWO-toptalent grant 021.001.040.

\bibliographystyle{apj}

\end{document}